\documentclass{revtex4-2}
\usepackage[utf8]{inputenc}
\usepackage{amsbsy}
\usepackage{amsopn}
\usepackage{amstext}
\usepackage{amsmath,amsthm,amsfonts,amssymb}
\usepackage[mathcal]{eucal}
\usepackage{mathrsfs}
\usepackage{braket}
\usepackage[english]{babel}
\usepackage{color}
\usepackage{esint}
\usepackage{graphicx}
\usepackage{float}
\usepackage{units}
\usepackage{textcomp}

\DeclareGraphicsExtensions{.png,PNG,.pdf,.PDF}
\usepackage{hyperref}
\usepackage{slashed}
\newcommand{\ie}{\begin{equation}}
\newcommand{\fe}{\end{equation}}
\newcommand{\se}{\begin{eqnarray}}
\newcommand{\ff}{\end{eqnarray}}
\begin{document}

\title{Comment on ``Deformations of the spin currents by topological screw dislocation and cosmic dispiration''}
\author{R. R. S. Oliveira}
\email{rubensrso@fisica.ufc.br}
\affiliation{Universidade Federal do Cear\'a (UFC), Departamento de F\'isica,\\ Campus do Pici, Fortaleza - CE, C.P. 6030, 60455-760 - Brazil.}


\date{\today}

\begin{abstract}
In this comment, we showed that the Dirac equation in the screw dislocation space-time also carries a term that represents the torsion of such topological defect, given by $K_\mu$. Therefore, the Dirac equation worked by Wang $et$ $al$. is incomplete since such a term was ignored in your equation (what cannot happen). In other words, it is only possible to work with the Dirac equation in the form presented by Wang $et$ $al$. if the space-time is torsion-free, which is obviously not the case.
\end{abstract}

\maketitle

\section{Introduction}

In a paper published in the Annals of Physics, Wang $et$ $al$. \cite{Wang} studied the spin currents induced by topological screw dislocation and cosmic inspiration via the Dirac equation in a curved space-time. By using the extended Drude model, the authors found that the spin-dependent forces are modified by the nontrivial geometry. For the topological screw dislocation, only the direction of the spin current was bended by deforming the spin polarization vector. In contrast, the force induced by cosmic dispiration could affect both the direction and magnitude of the spin current. As a consequence, the spin-Hall conductivity does not receive corrections from screw dislocation. In particular, this paper seems interesting and, in a way, brings innovative results.

However, investigating in detail some papers in the literature on the Dirac equation in the screw dislocation space-time, we verified that the Dirac equation worked by Wang $et$ $al$. \cite{Wang}, given by Eq. (1), is incomplete. Therefore, the objective of the present comment is to show that to work with the Dirac equation in the form presented by Wang $et$ $al$. \cite{Wang} is incoherent. In other words, we will see that it is only possible to work with the Dirac equation of Wang $et$ $al$. \cite{Wang} if the space-time is torsion-free (but this not is the case).


\section{The Dirac equation in the screw dislocation space-time}

Before we present the Dirac equation in the screw dislocation space-time, it is important to highlight that a dislocation is a type of linear topological defect (it can be of cosmic or condensed material origin) that carries torsion in its line element (or metric), arise from a break of the translational symmetry, and can be described by the Riemann–Cartan geometry \cite{da1,da2,da3,valanis,puntigam,katanaev}. In particular, in the case of the screw dislocations, they can manifest (or be classified) in two ways: the type-I screw dislocation, which is a distortion of a vertical line into a vertical spiral, and the type-II screw dislocation (Wang $et$ $al$. \cite{Wang} worked with this), which is a distortion of a circular curve into a vertical spiral (known as the Katanaev–Volovich dislocation) \cite{da1,da2,da3,valanis,puntigam,katanaev}.

In that way, the Dirac equation in the cosmic dislocation space-time must have a term that describes the torsion of such topological defect. So, according to the literature \cite{bakke0,bakke1,bakke2,bakke3,bakke4}, the $(3+1)$-dimensional covariant Dirac equation in the cosmic dislocation space-time (or any space-time with torsion) is given as follows (cylindrical coordinates with $\hbar=c=1$)
\begin{equation}\label{1}
\left[i\gamma^\mu(x)\nabla_\mu(x)-m\right]\Psi=0, \ \ (\mu=t,\rho,\varphi,z),
\end{equation}
where
\begin{equation}\label{2}
\nabla_\mu(x)=\partial_\mu(x)+\Gamma_\mu(x)+K_\mu(x),
\end{equation}
being $\nabla_\mu(x)$ the covariant derivative, $\partial_\mu$ are the usual partial derivatives, $\Gamma_\mu(x)=\frac{i}{4}\omega_{\mu ab}(x)\Sigma^{ab}$ is the spinorial connection, being $\omega_{\mu ab}$ the connection 1-form (spin connections) and $\Sigma^{ab}=\frac{i}{2}[\gamma^a,\gamma^b]$ is antisymmetric tensor, $K_\mu(x)=\frac{i}{4}K_{\mu ab}(x)\Sigma^{ab}$ is the contortion tensor, being $K_{\mu ab}(x)$ the ``torsion tensor'' (it actually depends on the torsion tensor), $\gamma^\mu=e^\mu_a (x)\gamma^a$ are the curved gamma matrices, being $e^\mu_a (x)$ the tetrads and $\gamma^a$ are the flat gamma matrices (or the standard gamma matrices in the Minkowski space-time), $m>0$ is the rest mass of the fermion, and $\Psi\in \mathbb{C}^4$ is the four-component Dirac spinor.

Now, with respect to the screw dislocation space-time, such background is modeled by the following line element with signature $(+,-,-,-)$ \cite{bakke0,bakke3,bakke4,valanis,puntigam,katanaev}
\begin{equation}
ds^2=dt^2-d\rho^2-\rho^2 d\varphi^2-(dz+\chi d\varphi)^2,
\end{equation}
where $\chi$ is a constant real parameter that is related to the torsion (field) of the defect, or, by using the crystallography language, is related to the Burgers vector $\Vec{b}=b \Vec{e}_z$ ($\chi=b/2\pi$).

Therefore, as the Dirac equation \eqref{1} in the screw dislocation space-time also carries a term that represents the torsion of the defect (given by $K_\mu$), implies that the Dirac equation
presented by Wang $et$ $al$. \cite{Wang}, given by Eq. (1), is incomplete, i.e., the term $K_\mu$ was ignored or forgot (what cannot happen). In other words, Wang $et$ $al$. \cite{Wang} worked with a Dirac equation that is only valid (permitted) when the space-time is torsion-free, such as occurs in Refs. \cite{B1,B2,oliveira1,oliveira2,oliveira3,oliveira4}. Besides, the peculiar thing about all this is that Wang $et$ $al$. \cite{Wang} goes so far as to state in your article that the parameter $\chi$ is the torsion of the screw dislocation (what is truth), and also cites some papers with the correct Dirac equation, such as Refs. \cite{bakke3,bakke4}; however, even so, he ignored (or forgot) the term $K_\mu$.


\section{Final remarks}

In this comment, we showed that the Dirac equation in the screw dislocation space-time also carries a term that represents the torsion of the defect, given by $K_\mu$ \cite{bakke0,bakke1,bakke2,bakke3,bakke4}. Therefore, the Dirac equation presented by Wang $et$ $al$. \cite{Wang} is incomplete since the term $K_\mu$ was ignored (or forgot) in your Dirac equation (what cannot happen). In other words, it is only possible to work with the Dirac equation in the form presented by Wang $et$ $al$. \cite{Wang} if the space-time is torsion-free, which is obviously not the case.

\section*{Acknowledgments}

\hspace{0.5cm}The author would like to thank the Conselho Nacional de Desenvolvimento Cient\'{\i}fico e Tecnol\'{o}gico (CNPq) for financial support.

\section*{Data availability statement}

\hspace{0.5cm} This manuscript has no associated data or the data will not be deposited. [Author’ comment: There is no data associated with this manuscript or no data has been used to prepare it.]

\end{document}